\documentclass[final,3p,times,twocolumn]{elsarticle}
\usepackage{tabularx}
\usepackage{lineno}
\usepackage{graphics}
\usepackage{graphicx}
\usepackage{amssymb}
\usepackage{amsthm}
\usepackage{amsmath}
\usepackage{upgreek}
\usepackage{subfigure}
\usepackage{wasysym}
\usepackage{textcomp}
\usepackage{array}
\usepackage{booktabs}
\usepackage{paralist}
\usepackage{verbatim}
\usepackage{float}

\biboptions{square}
\journal {Nuclear Instruments and Methods in Physics Research Section A}

\begin{document}
%\linenumbers
\begin{frontmatter}

\title{MWPC prototyping and performance test for the STAR inner TPC upgrade}

\cortext[cauthor]{Correspondence to: 27 Shanda South Road, Jinan 250100, Shandong, China.\\
E-mail address: xuqh@sdu.edu.cn, chiyang@sdu.edu.cn}

\author[1:sdu]{Fuwang Shen}  %Fuwang Shen
\author[1:sdu]{Shuai Wang}  %Shuai Wang
\author[1:sdu]{Fangang Kong} %Fangang Kong
\author[1:sdu]{Shiwei Bai} %Shiwei Bai
\author[1:sdu]{Changyu Li}   %Changyu Li
\author[2:STAR]{Flemming Videb{\ae}k} %Flemming Videbaek
\author[1:sdu,2:STAR]{Zhangbu Xu} %Zhangbu Xu
\author[1:sdu]{Chengguang Zhu}  %Chengguang Zhu
\author[1:sdu]{Qinghua Xu\corref{cauthor}}   %Qinghua Xu
\author[1:sdu]{Chi Yang\corref{cauthor}}  %Chi Yang

\address[1:sdu]{School of Physics and Key Laboratory of Particle Physics and Particle Irradiation (MOE), Shandong University, Jinan 250100, China}
\address[2:STAR]{Department of Physics, Brookhaven National Laboratory, Upton, NY 11973, USA}

\begin{abstract}
A new prototype of STAR inner Time Projection Chamber (iTPC) MWPC sector has been fabricated and tested in an X-ray test system. The wire chamber built at Shandong University has a wire tension precision better than 6$\%$ and wire pitch precision better than 10 $\mu$m. The gas gain uniformity and energy resolution are measured to be better than 1$\%$ (RMS) and 20$\%$ (FWHM), respectively, using an $^{55}$Fe X-ray source. The iTPC upgrade project is to replace all 24 STAR TPC inner sectors as a crucial detector upgrade for the RHIC beam energy scan phase II program. The test results show that the constructed iTPC prototype meets all project requirements.
\end{abstract}

\begin{keyword}
STAR; TPC; MWPC; X-ray test system; Gain uniformity
\end{keyword}

\end{frontmatter}

\section{Introduction}

The STAR experiment is upgrading the inner sectors of Time Projection Chamber (iTPC) detector to re-instrument the inner pad planes and renew the inner sector wire chambers \cite{ref:det1,ref:det2,ref:note}. The upgrade will expand the TPC acceptance from pseudo-rapidity $|$$\eta$$|$$\leq$1.0 to $|$$\eta$$|$$\leq$1.5, and enable better acceptance for tracks with low momentum, as well as better resolution in both momentum and energy loss $dE/dx$ for tracks of all momenta. The enhanced capabilities achieved from the iTPC upgrade are crucial to the physics program of the Beam Energy Scan II (BES-II) \cite{ref:note2} at RHIC during 2019$\sim$2020, in particular the study of the QCD phase transition and the search for a possible QCD critical point.

The TPC is the primary tracking device at STAR, which records the tracks of particles, measures their momenta in a 0.5 T magnetic field, and identifies the particles by measuring their ionization energy loss $dE/dx$ \cite{ref:a,ref:b}. The two endcaps of the TPC consists of 24 inner trapezoidal Multi Wire Proportional Chamber (MWPC) sectors, where the drift electrons are collected and avalanched. The original pad planes of the STAR TPC inner sectors are sparsely instrumented with 13 pad-row readouts at a level of 20$\%$ areal population. The iTPC upgrade project will replace all 24 inner sectors in the STAR TPC with new, fully instrumented, sectors with 40 pad-row readouts. The number of pads is increased from 1750 to 3440, with an enlarged pad size. The upgrade requires the new MWPC wire grids to operate at lower gain and to utilize larger pads. A set of corresponding new iTPC electronics are being designed to handle the factor-of-two increased number of readout channels \cite{ref:note}. It includes preamplifier and digitizer ASIC ("SAMPA", designed for ALICE TPC upgrade), front end electronics cards ("iFEE"), readout boards ("iRDO") and data acquisition system (DAQ).

\begin{figure*}[hbtp]
\begin{center}
\includegraphics[width=0.9\textwidth]{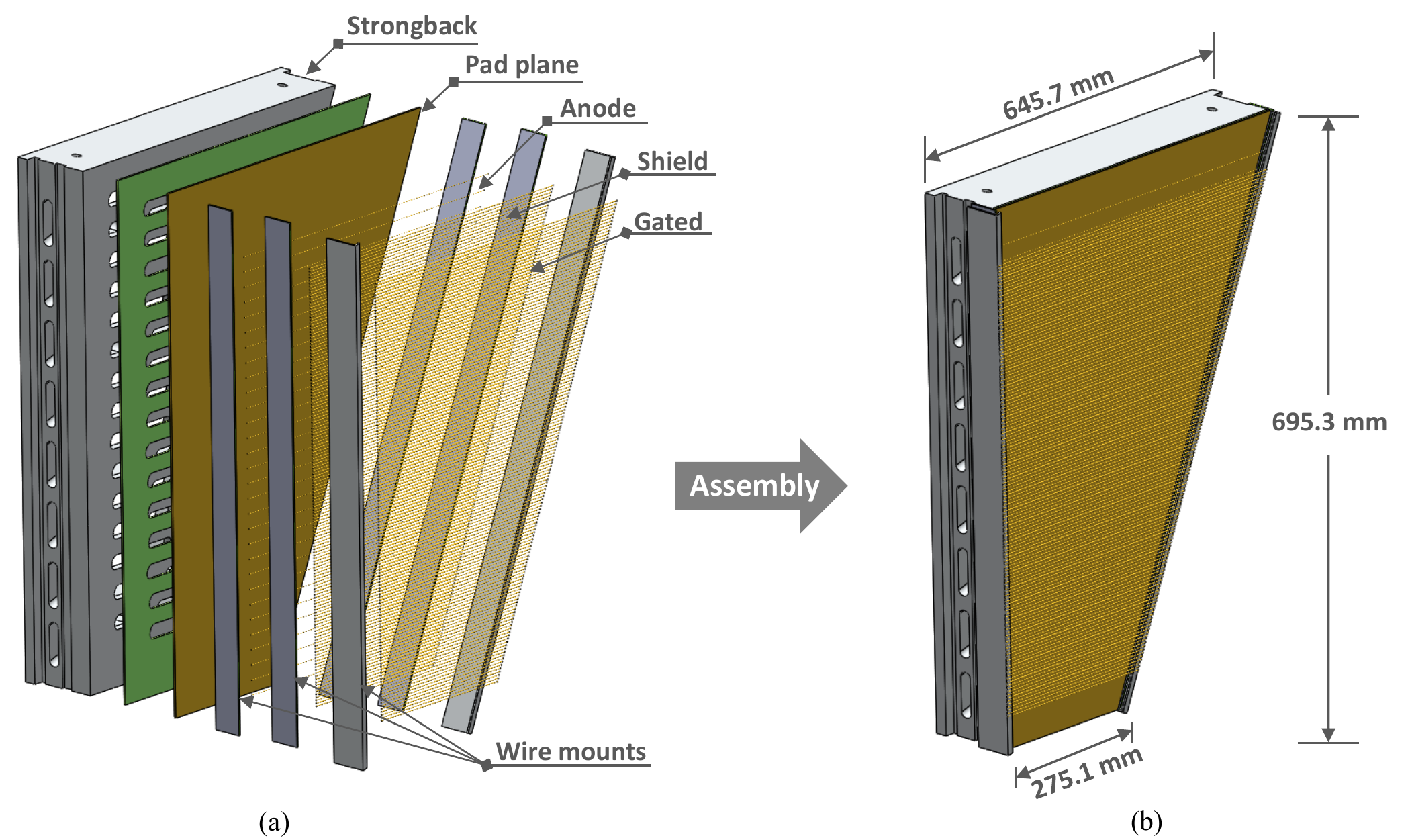}
\caption{(a) The exploded view of the STAR iTPC sector. Each sector consists of strongback, PCB pad plane, three layers of wires and the corresponding wire mounts. (b) The assembly of the iTPC sector.}
\label{fig:MWPC}
\end{center}
\end{figure*}

The structure of the iTPC MWPC sector is shown in Figure~\ref{fig:MWPC}, which includes Aluminium supporting strongback, PCB pad plane board, and three layers of wires. High precision of wire location and wire tension of the wire grids on Aluminum strongback are required during the assembly of the MWPC sectors in order to meet the desired performance, which includes excellent gas gain uniformity, energy resolution and stability. A full size iTPC MWPC prototype had been assembled and tested at Shandong University (SDU). The details of iTPC prototype construction and performance are presented in this paper.

\section{iTPC prototype sector construction}

The iTPC sector construction includes several steps: pad plane bonding, wire mounts installation and pinning, wires winding, wire plane mounting, epoxy and soldering of the wires. All these steps require strict control of the mechanical tolerance of the components such as wire pitch, wire tension and the distance between pad plane and wire plane. These requirements will provide the geometry uniformity in drifting and ionization zones. The characteristics of the wires are shown in the Table\ref{Wires}. The details for both wire winding and mounting are presented in the subsections.

\begin{table}[h]\small
\centering
\begin{center}
\begin{tabular}{m{3cm}m{1cm}m{1cm}m{1cm}}
\hline\specialrule{0em}{2pt}{2pt}
Wire                    & Anode & Shield & Gated\\\specialrule{0em}{2pt}{2pt}
\hline\specialrule{0em}{1pt}{1pt}
Material                & W     & Be-Cu  & Be-Cu\\\specialrule{0em}{1pt}{1pt}
Diameter ($\mu$m)       & 20    & 75     & 75\\\specialrule{0em}{1pt}{1pt}
Pitch (mm)              & 4     & 1      & 1 \\\specialrule{0em}{1pt}{1pt}
Dist. to pad plane (mm) & 2     & 4      & 10 \\\specialrule{0em}{1pt}{1pt}
Tension (N)             & 0.5   & 1.2    & 1.2\\\specialrule{0em}{1pt}{1pt}
\hline
\end{tabular}
\end{center}
\caption{The iTPC MWPC wires parameters}
\label{Wires}
\end{table}

\subsection{wire winding}

There are 164 anode wires, 681 shield wires and 681 gated wires for each iTPC sector. Before mounted on the sector, all the wires are first wound on wire transfer frames using a wire winding machine, which consists of a swiveling table, incremental encoder, step motors, tension sensor, pulley system and the motion controller. The transfer frames are mounted on the swiveling table. The step motor and incremental encoder control the rotation of the table and the step size or the wire pitch. The tension sensor monitors the wire tension and provides feed back to the motion controller to make sure the wire tension fall in required range.

\begin{figure*}
\begin{center}
\includegraphics[width=1.0\textwidth]{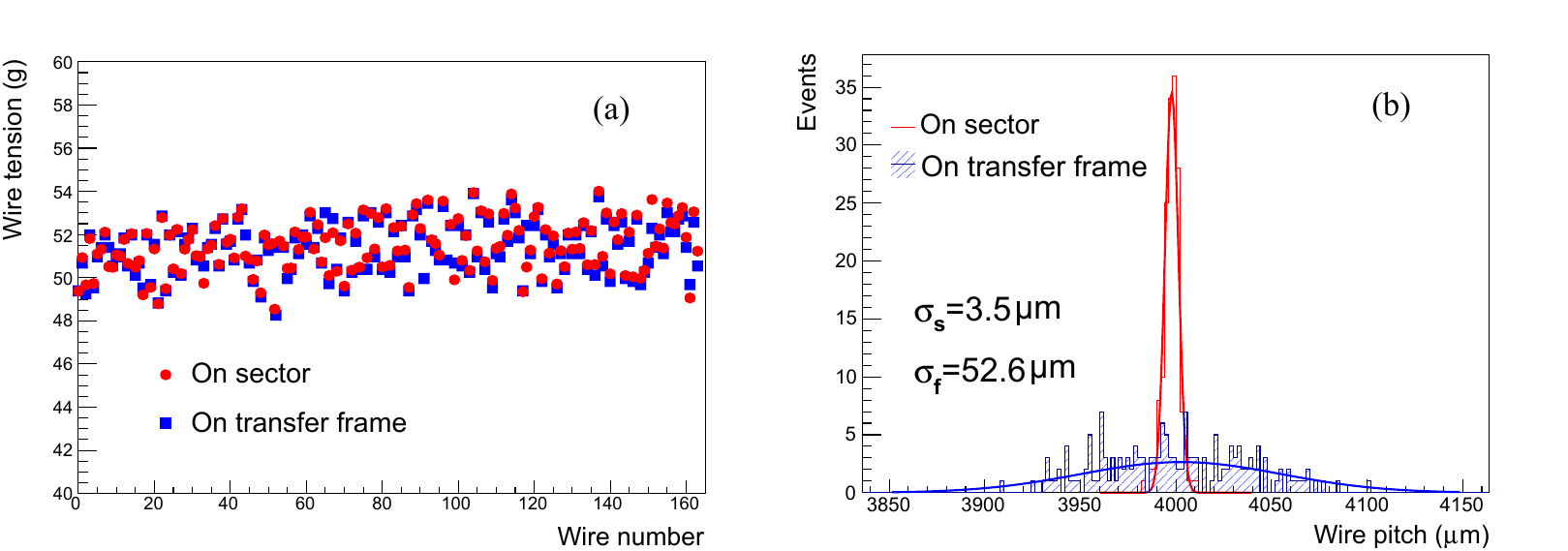}
\caption{(Color online) (a) The anode wire tension on sector and on wire transfer frame, all data falls within the required range of 51$\pm$3 g. (b) The anode wire pitch distribution on sector and on wire transfer frame. Gaussian functions are used to fit the data. The $\sigma_s$ and $\sigma_f$ are the standard deviation of wire pitch distribution on sector and on wire transfer frame, respectively.}
\label{fig:Tension}
\end{center}
\end{figure*}

To inspect the wire tension and pitch, a laser based wire tension measurement system has been designed and implemented. Basically, the wire tension is determined by measuring the fundamental vibration frequency from air-jet induced vibration with a diode laser based optical system. The details can be found in Ref. \cite{ref:c}. Figure~\ref{fig:Tension} (a) shows the wire tension measurement results for anode wires while on transfer frame and after being mounted on sector, and the tension falls within the required range of 51$\pm$3 g (0.5$\pm$0.03N).

\subsection{wire mounting}

The wire plane is mounted on the wire mounts after the wires tension inspection. The transfer frame is placed on the wire mounting system, which consists of a granite table with good flatness, lead rail and wire combs. The lead rail allows us to adjust the wire transfer frame position to mount the wires into the comb. Wires are then glued and soldered. The comb serves to control precisely the wire pitch and height \cite{ref:d}. The wire pitch can be scanned by the grating ruler of the laser based system as mentioned above. As shown in Figure~\ref{fig:Tension} (b), the wire pitch are improved significantly after using wire comb and a position precision better than 10 $\mu$m is achieved. Then layer of wires are mounted and soldered, respectively, for each wire plane.
\begin{figure*}
\begin{center}
\includegraphics[width=0.8\textwidth]{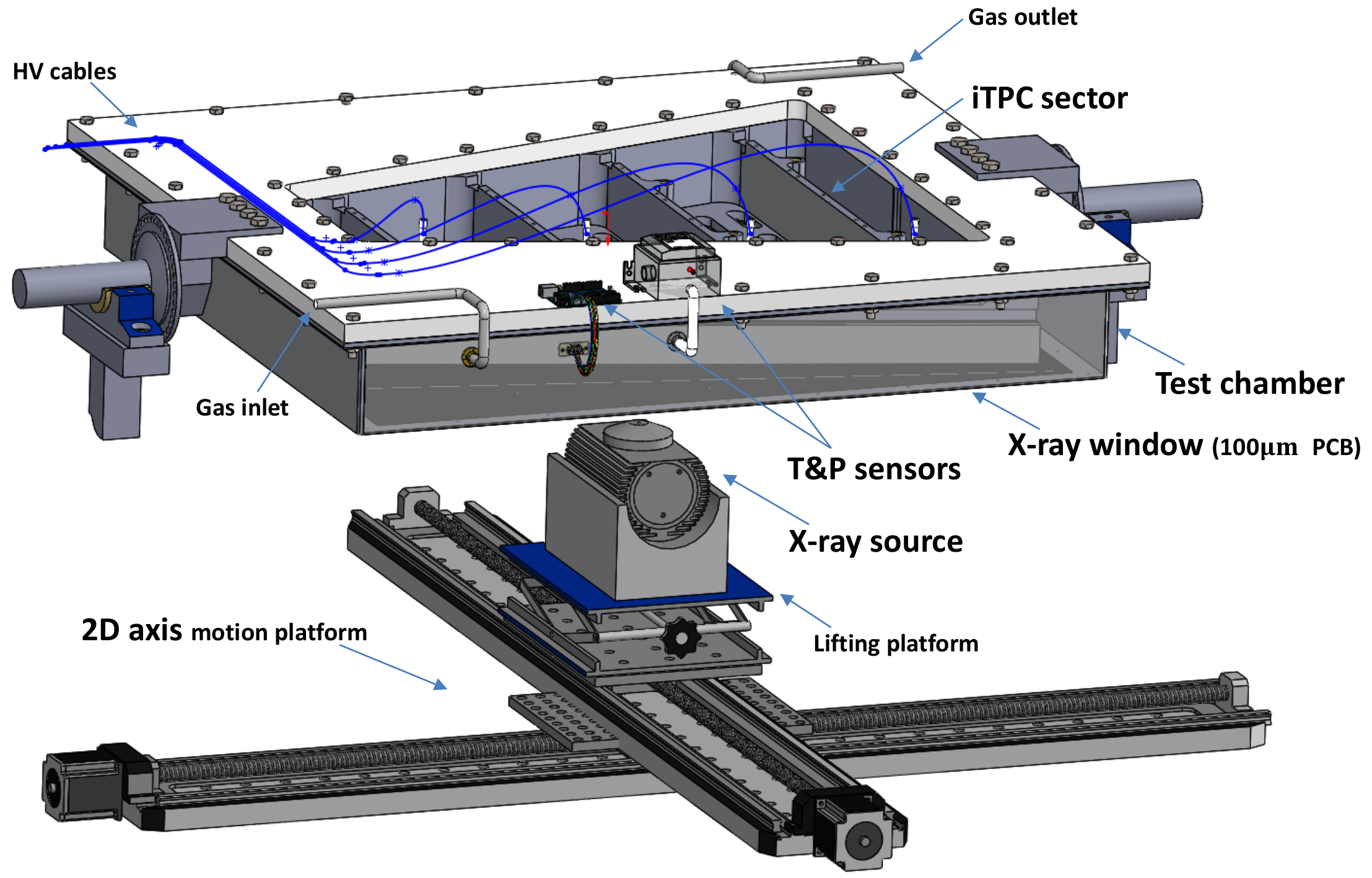}
\caption{The layout of the X-ray test system. It consists of a test chamber, 2D axis stepper platform, X-ray source, temperature and pressure sensors.}
\label{fig:TestSystem}
\end{center}
\end{figure*}

\section{X-ray testing system}

After all the construction steps, a full size iTPC prototype sector has been made. To test the performance of the whole sector , an X-ray test system is designed and implemented. As shown in Figure~\ref{fig:TestSystem} this system includes  a steel supporting frame, a test chamber, a 2D axis stepper platform, X-ray source, temperature and pressure sensors. For the test chamber, the X-ray window is a 0.1 mm thick carbon coated PCB board, which can provide a short drift field for cosmic ray tests. The 2D axis stepper platform supports and moves the X-ray source (a 0.1 mCi $^{55}$Fe, or a 50 kV, 1000 $\mu$A Cu target X-ray tube). A gas mixer with a 0.1$\%$ mixing precision is used. The MWPC sector is installed in the test chamber and and flushed with P10 (90$\%$Ar + 10$\%$CH$_4$) gas. The pressure of the gas is set to 2 mbar above the atmosphere. The distance from the X-ray window to the anode wires is $10$ mm. The diameter of the $^{55}$Fe source is 5 mm (not collimated), and its X-rays covers up to 40 anode wires.  To readout the anode wire signal, a charge sensitive preamplifier and shaper Amptek A225 and a Multichannel analyzer (MCA) Amptek 8000D (13 bit ADC) is used.

\begin{figure}
\begin{center}
\includegraphics[width=0.45\textwidth]{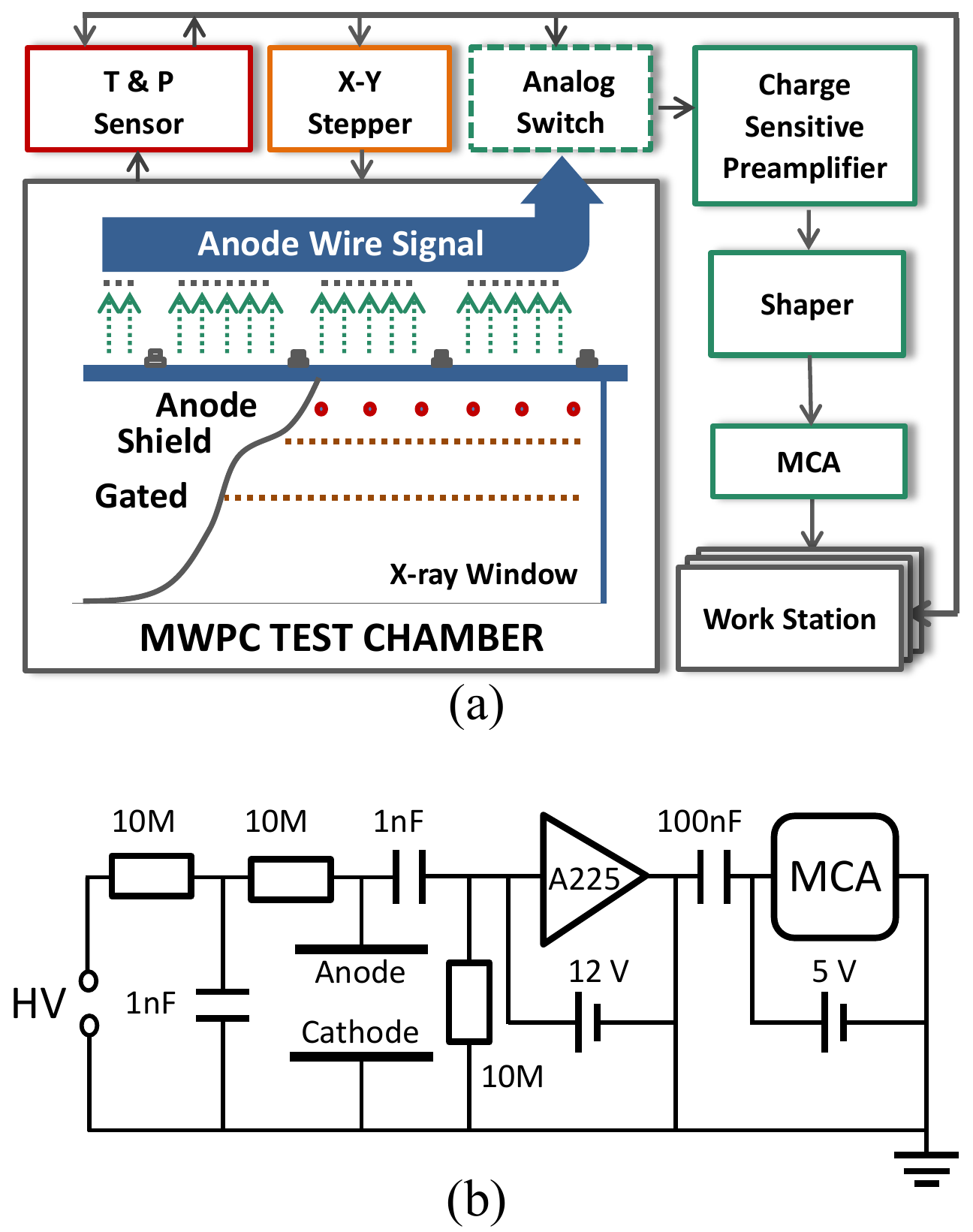}
\caption{(a) The MWPC performance test flow.  (b) The layout of the test system circuit.}
\label{fig:TestFlow}
\end{center}
\end{figure}

Figure~\ref{fig:TestFlow} (a) shows the performance test flow. The sensors monitor the gas temperature and pressure in the chamber every 5 seconds. The work station controls the stepper system to locate the source under specified wire or position, and then the preamplifier and MCA are used to readout the signals from anode wires. All the data acquired by the system is stored in work station storage for the further data analysis and backups.

\begin{figure*}
\begin{center}
\includegraphics[width=1.0\textwidth]{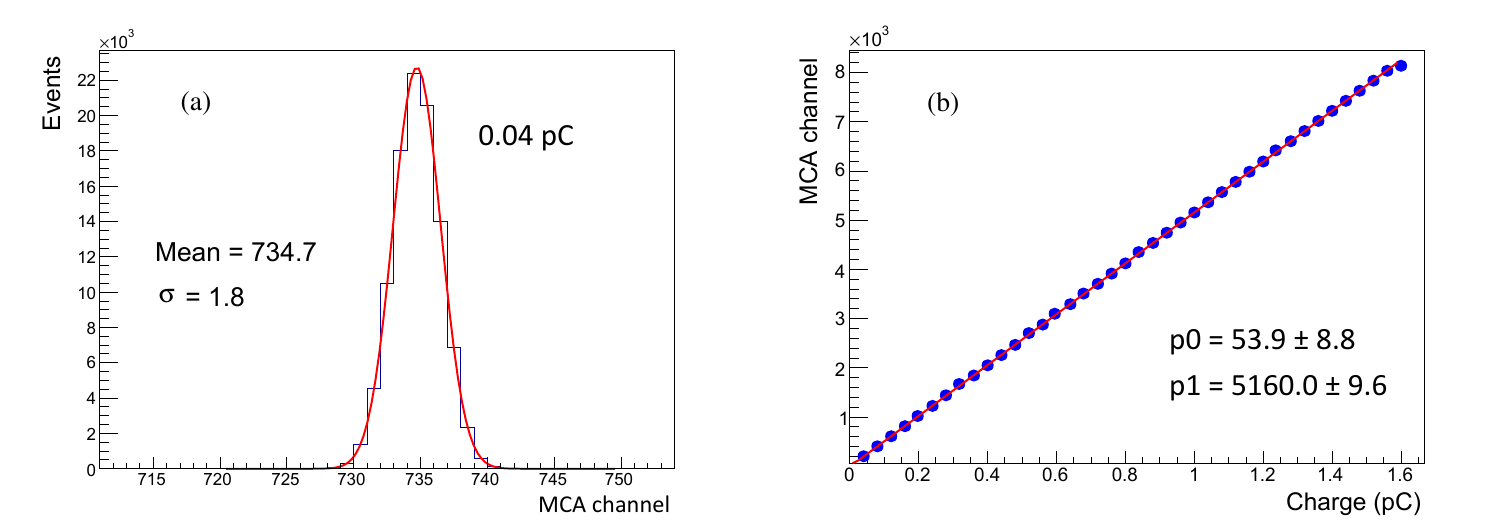}
\caption{(a) The MCA channel distribution from a input charge of 0.04 pC. (b) The average MCA channel versus different charge inputs, showing a good linearity in the working range of MCA.}
\label{fig:calibration}
\end{center}
\end{figure*}

Figure~\ref{fig:TestFlow} (b) shows the system circuit, which has a 2.4 $\mu$s peaking time. The MCA and preamplifiers with this circuit were calibrated by signals from a precise capacitor injected with a known amount of charge.  Figure~\ref{fig:calibration} shows preamplifiers and MCA calibration results. The charge resolution is less than 0.6$\%$ and good linearity is obtained between MCA channel and different input charge.

\section{iTPC prototype performance test}

The iTPC prototype has been tested with the $^{55}$Fe source and the Cu target X-ray tube. The gas gain dependence on high voltage, X-ray count rate and $T/P$ has been studied. The gas gain and energy resolution uniformity for anode wires has been scanned and the irradiation test also has been done. Except for the irradiation test, all the tests are done with the $^{55}$Fe source.

\subsection{X-ray spectrum}

\begin{figure}
\begin{center}
\includegraphics[width=0.45\textwidth]{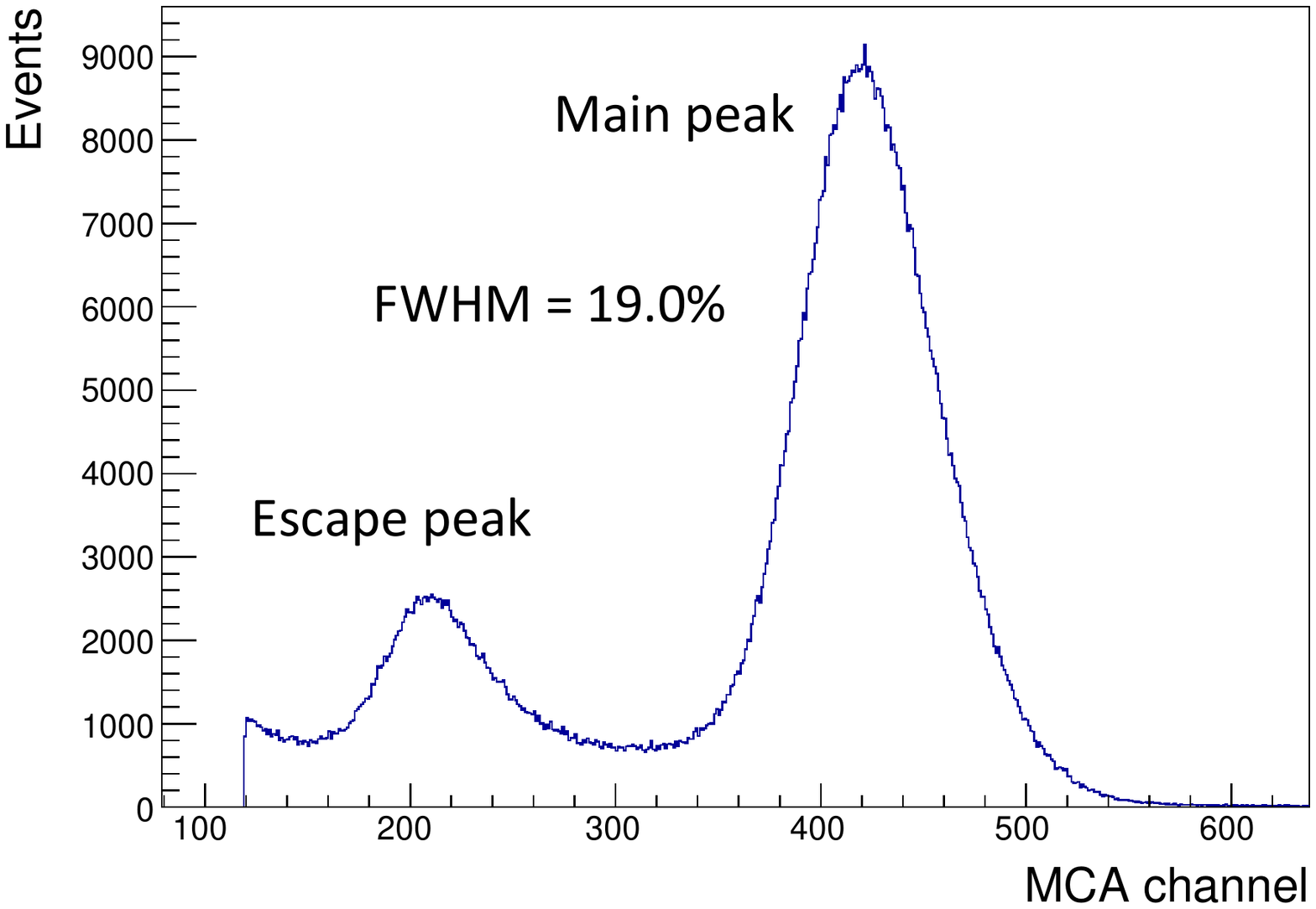}
\caption{The $^{55}$Fe X-ray spectrum of single anode wire from MCA with 1120 V anode voltage.}
\label{fig:Xspectrum}
\end{center}
\end{figure}

The $^{55}$Fe source decays via electron capture, and mainly emits Mn K-shell 5.9 keV X-rays. The 5.9 keV X-ray produces approximately 225 electrons for main peak and 110 electrons for Ar escape peak before avalanche in argon base detector. Figure~\ref{fig:Xspectrum} shows the single anode wire X-ray spectrum from MCA with the anode working voltage 1120 V. It shows clearly the corresponding two peaks. For the $^{55}$Fe X-ray spectrum, the peak position ratio of main peak to escape peak is about 2.0 and the energy resolution is 19$\%$ (FWHM).

\subsection{Gas gain dependence on anode voltage}

\begin{figure}
\begin{center}
\includegraphics[width=0.43\textwidth]{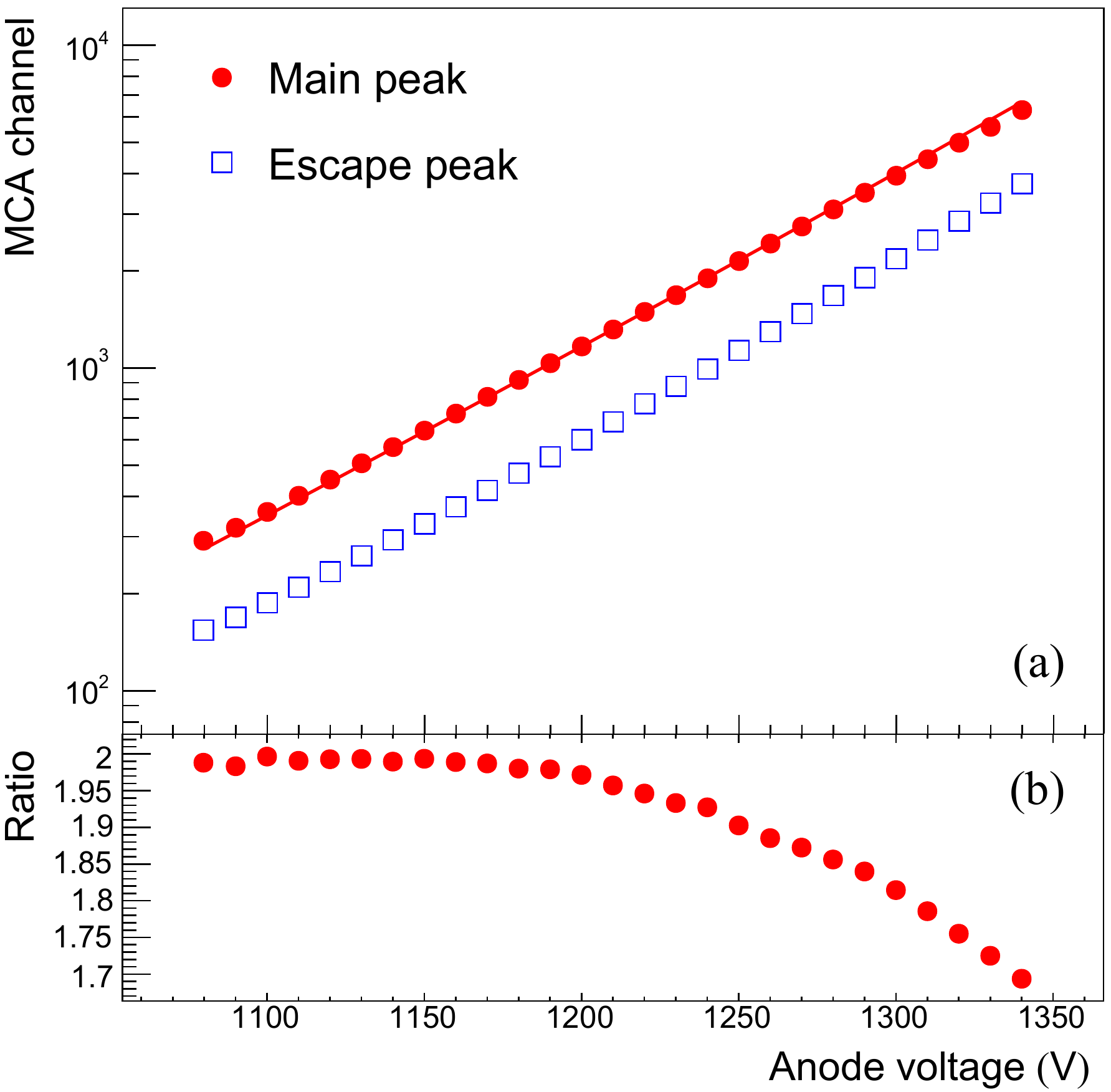}
\caption{(Color online) (a) The MCA channels versus anode voltage. Red line is the gain prediction of Diethorn formula (Eq.~\ref{eq:GainV}). (b) The peak position ratio of main peak to escape peak. As a self-induced space charge effect, the ratio decreases as the high voltage increases.}
\label{fig:GainVSVoltage}
\end{center}
\end{figure}

A high-voltage scan (1080 V to 1340 V) of anode wire with fixed gas condition has been done. The gas gain exponentially increases with the high voltage as shown in the Figure~\ref{fig:GainVSVoltage} (a). The gas gain $G$ is often described with the Diethorn formula \cite{ref:f}:

\begin{equation}
lnG = \frac{CVln2}{2\pi\varepsilon_0\Delta V}ln\frac{CV}{2\pi\varepsilon_0r_aE_{min}(\rho/\rho_0)},
\label{eq:GainV}
\end{equation}
where $V$ is the anode voltage, $\varepsilon_0$ is the permittivity of vacuum, $C$ is the capacitance per unit length of anode wire, $E_{min}$ is the critical electric field to avalanche, $\Delta V$ is the average energy needed to produce a secondary electron, $\rho$$/$$\rho_0$ is the gas density ratio between real-time and normal condition.  $E_{min}$ and $\Delta V$ are gas-dependent and voltage-independent parameters, which can be calculated by the measurement results of voltage scan \cite{ref:g}. The Diethorn parameters obtained here are: $E_{min} = 44.2\pm 0.5$ kV/cm, $\Delta V = 23.5\pm 0.2$ V. Figure~\ref{fig:GainVSVoltage} (b) shows the peak position ratio of main peak to escape peak. It decreases from 2.0 to 1.7 as the high voltage increases. This is caused by the self-induced space charge effect on gas gain \cite{ref:h}. The generated electrons in avalanche produce the space charge and effective voltage drop between anode and cathode. The gas gain obtained by main peak (5.9 keV) primary electron becomes increasingly lower than that obtained by the escape peak (2.9 keV) primary electron as the anode voltage increases.
\subsection{Gas gain dependence on count rate}

\begin{figure}
\begin{center}
\includegraphics[width=0.45\textwidth]{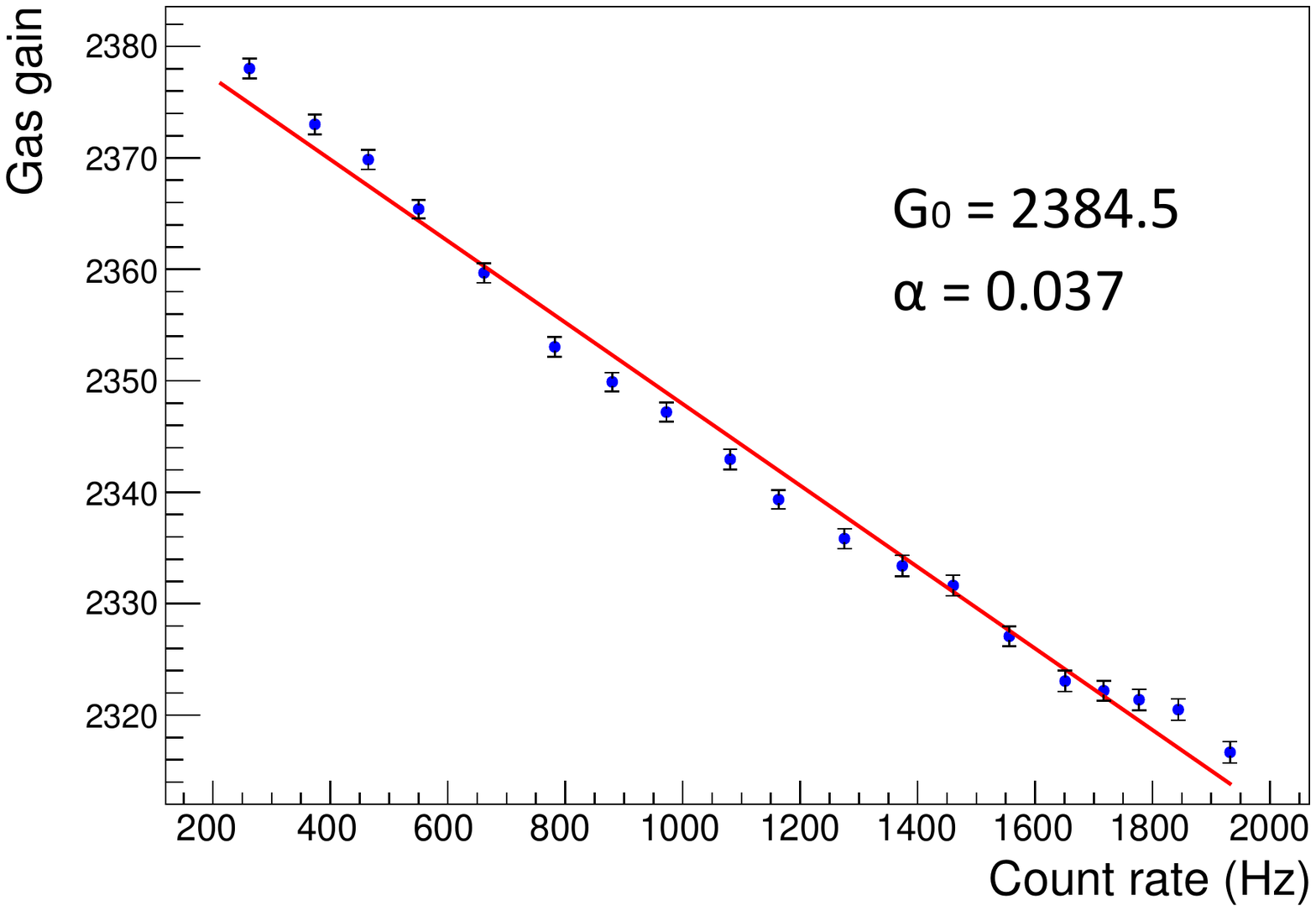}
\caption{(Color online) The gas gain versus X-ray count rate. It is fitted with a linear function and the corresponding parameters are shown.}
\label{fig:GainEventRate}
\end{center}
\end{figure}

The $^{55}$Fe source is not collimated and its X-rays covers up to 40 anode wires as mentioned above. For the anode wire readout with $^{55}$Fe source, the count rate varies with the anode wires as the distance from wire to $^{55}$Fe source is different. The gas gain dependence on count rate is studied. The maximal count rate of X-ray detected by single anode wire with a fixed MCA threshold is about 2000 Hz. As shown in Figure~\ref{fig:GainEventRate}, the gas gain decreases as the X-ray count rate increases. This effect had been researched by several authors. R.W. Hendricks found the space charge effects in proportional counters \cite{ref:i} and Ken Katagiri had developed a numerical model to calculate the space charge effect in MWPC \cite{ref:j}. W. Riegler summarized some typical models of space charge effect in wire chambers \cite{ref:k,ref:l,ref:m}. The effect of the space charge with simple spatial distribution on the electric field in the detector can be found by solving Green function and Poisson equation. The effective voltage drop $\delta V$ between anode and cathode is proportional to the space charge density $\rho_e$. $\rho_e$ is proportional to the X-ray count rate $R$. Since $\delta V$ is very small relative to the anode voltage $V$. The gas gain can be described by

\begin{equation}
G = {G}_0 - \alpha R.
\end{equation}

As shown in Figure~\ref{fig:GainEventRate}, the measured gain indeed decrease linearly versus count rate. As the $^{55}$Fe X-ray count rate increases from 200 Hz to 2000 Hz, a gas gain decrease of about 2.7$\%$ is observed.

\subsection{Gas gain dependence on T/P}

We have studied the gas gain variation effected by temperature and pressure. For a small fluctuation of the gas condition, the gas gain is  inversely proportional to the gas density \cite{ref:g,ref:n}. Assuming the P10 behaves as ideal gas, the corresponding change of gas gain versus the variation of $T/P$ can be described by

\begin{equation}
\frac{\Delta G}{G} = \beta\frac{\Delta (T/P)}{T/P} + b.
\end{equation}

\begin{figure*}
\begin{center}
\includegraphics[width=1.0\textwidth]{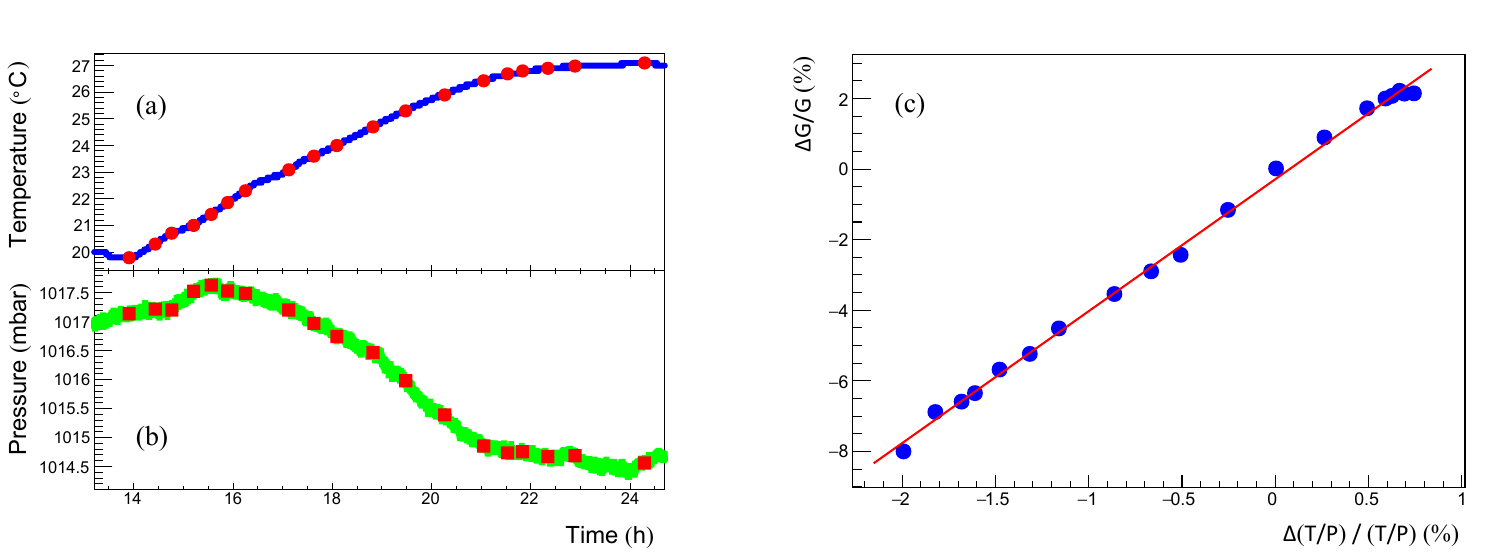}
\caption{(Color online) (a) The real-time temperature and (b) the pressure inside the test chamber during $\sim$10 hours' test. The red points represent the corresponding values while reading out signal from the anode wire. (c) The corresponding gas gain fluctuation versus $T/P$ variation. The point at zero represents the data under 25 $^\circ$C and 1015 mbar}
\label{fig:Temp}
\end{center}
\end{figure*}

The increase of $T/P$ reduces the gas number density, and the gas gain increase  proportionally as $T/P$ increases. To study the relation between gas gain and $T/P$, the temperature and pressure sensors acquired the $T/P$ data inside test chamber every 5 seconds, as shown in Figure~\ref{fig:Temp} (a) and (b). In the ten hours test, the temperature inside chamber varied from 20 $^\circ$C to 27 $^\circ$C, and the pressure varied from 1014 mbar to 1017 mbar. The gain varied about 10$\%$ as shown in Figure~\ref{fig:Temp} (c). The standard gain is taken under 25 $^\circ$C and 1015 mbar. From the fitting function of the measurement results, the $\beta = 3.74\pm 0.04$, $b = -0.29\pm 0.04$. The effect of $T/P$ is important to correct the gas gain under different $T/P$ condition while scanning all anode wires.

\subsection{Gas gain and energy resolution uniformity}

\begin{figure}
\begin{center}
\includegraphics[width=0.45\textwidth]{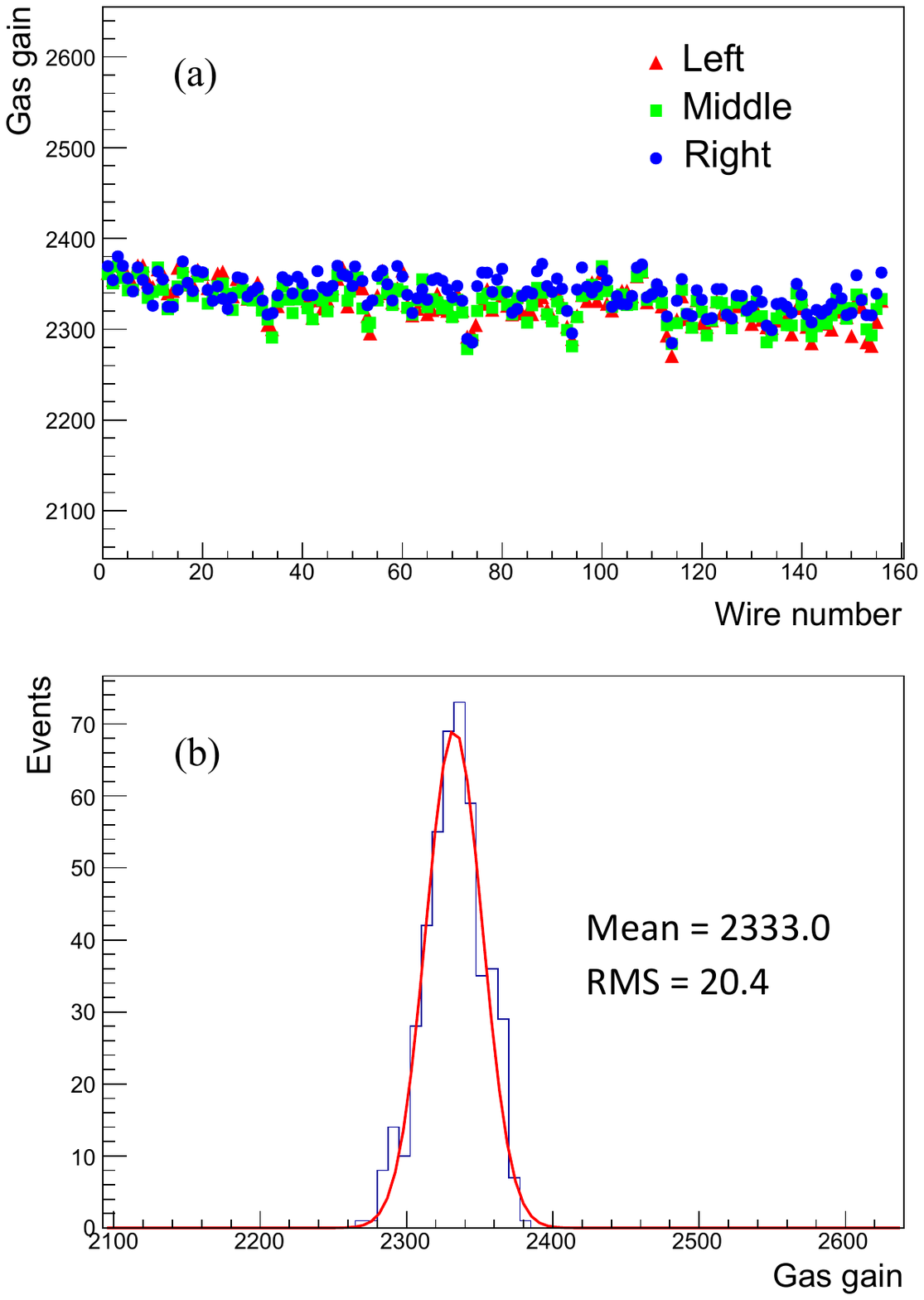}
\caption{(Color online) (a) The gas gain  on $3$ different positions (left, middle, right) of anode wires. (b) The gas gain distribution of anode wires.}
\label{fig:GainUni}
\end{center}
\end{figure}

\begin{figure}
\begin{center}
\includegraphics[width=0.45\textwidth]{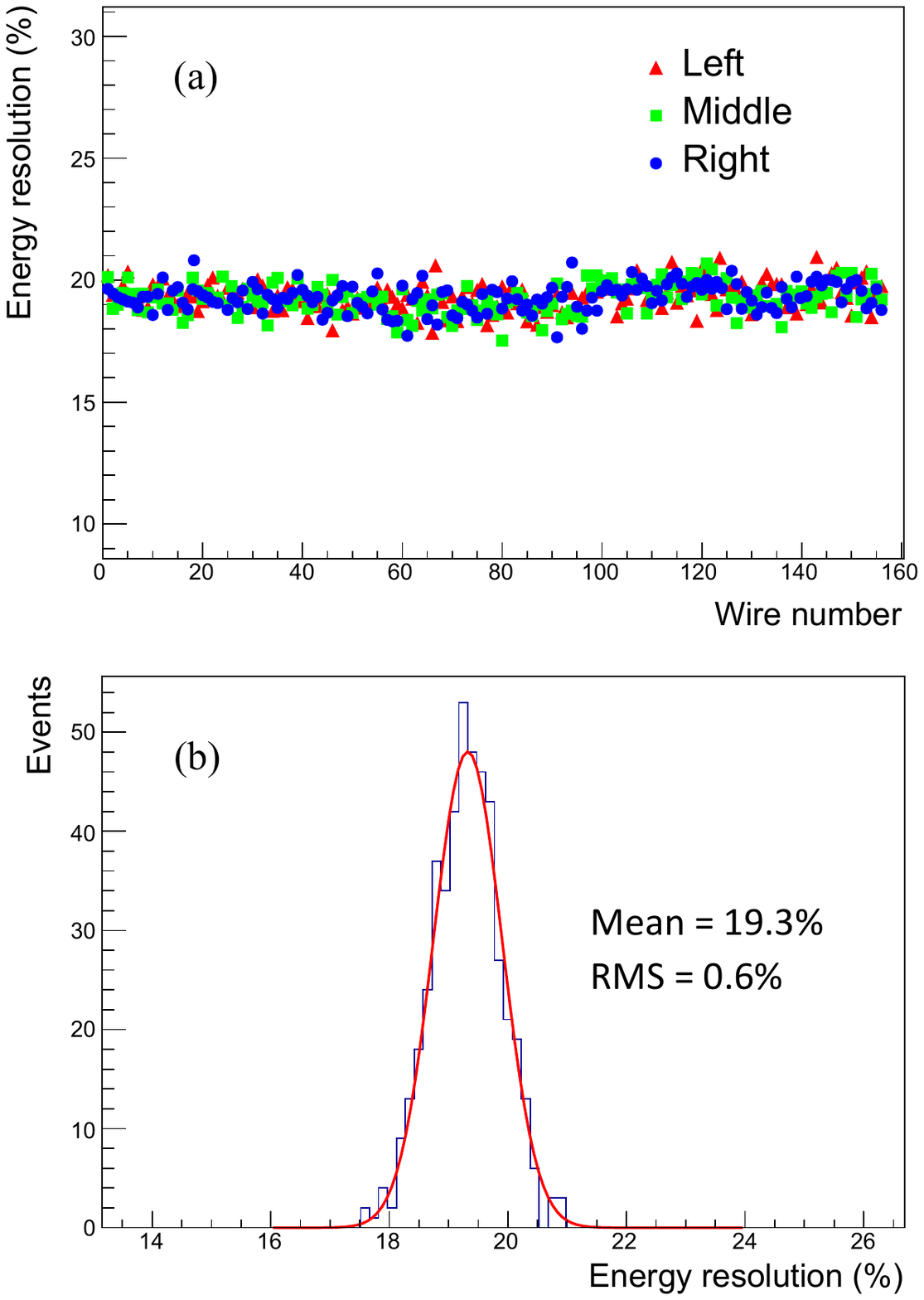}
\caption{(Color online) (a) The energy resolution (FWHM)  on $3$ different positions (left, middle, right) of  anode wires. (b) The energy resolution (FWHM) distribution of anode wires.}
\label{fig:ERWire}
\end{center}
\end{figure}

For the gas gain uniformity, all the anode wires at 3 different positions for each wire have been scanned with the $^{55}$Fe source. The source was fastened to the 2D stepper system, which moved the source to the specified wire location. The scan was done wire by wire, and due to the readout channel numbers limit, 156 anode wires are available to be readout. Figure~\ref{fig:GainUni} (a) shows the gas gain versus wire number. Figure~\ref{fig:GainUni} (b) shows the gas gain distribution of all available anode wires, which has corrected the effect of the X-ray count rate and $T/P$. The gas gain uniformity $RMS/Mean$ is about 0.9$\%$, which is better than the 2$\%$ requirement. Figure~\ref{fig:ERWire} (a) shows the energy resolution versus wire number, and Figure~\ref{fig:ERWire} (b) shows the energy resolution distribution of all available anode wires. The energy resolution (FWHM) is about 19.3$\%$, which is also better than the 24$\%$ requirement at STAR.

\subsection{Irradiation test}

\begin{figure}
\begin{center}
\includegraphics[width=0.45\textwidth]{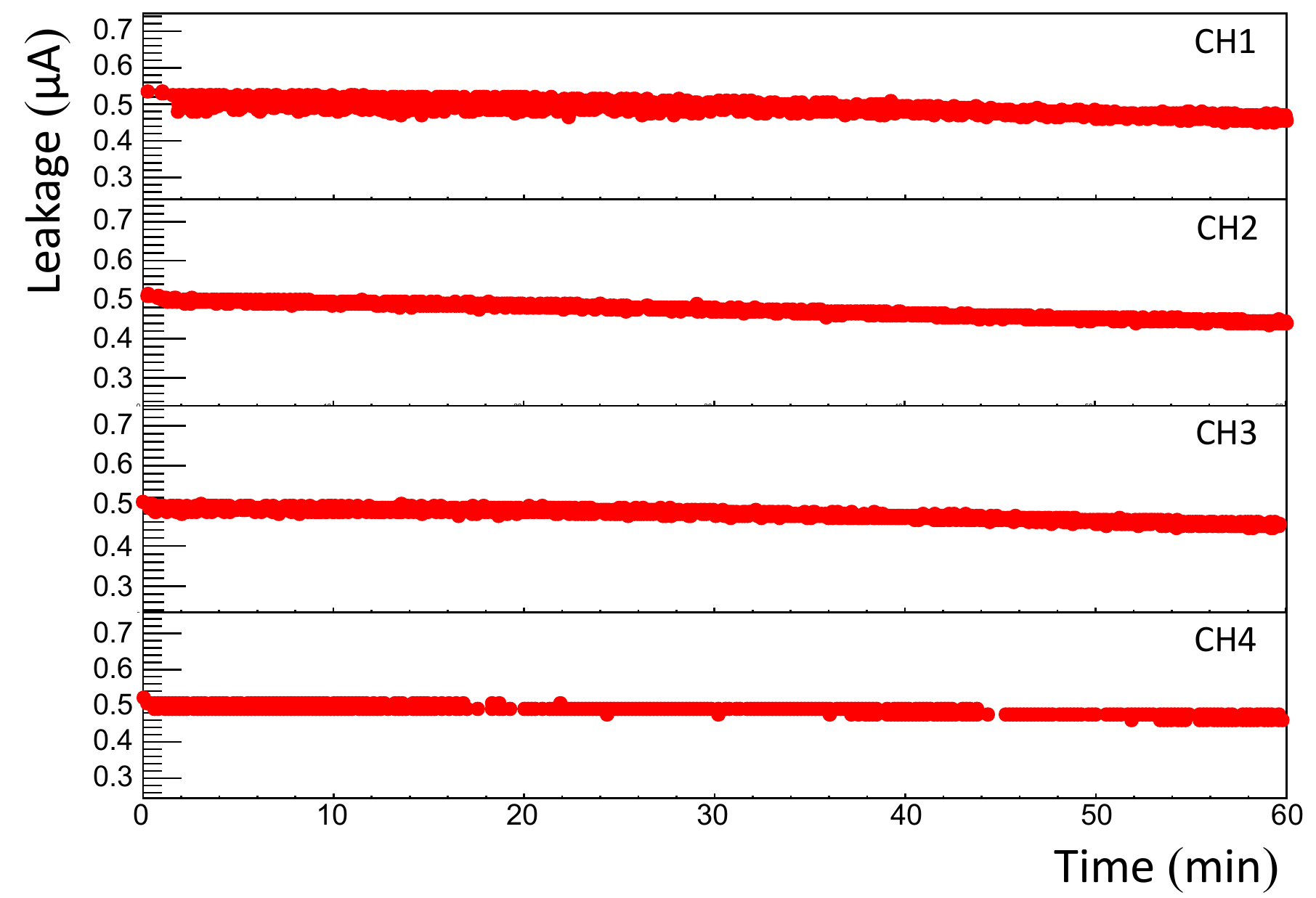}
\caption{The MWPC leakage current of four HV channels versus time. The leakage current reaches 500 nA without trip or any sparks.}
\label{fig:StrongTest}
\end{center}
\end{figure}

To test the stability of the whole sector, the irradiation test was done with the Cu target X-ray tube. Adjusting the tube voltage and current, the energy of the X-rays could vary from 0 keV to 50 keV, and the count rate can be as high as 230 kHz. The X-ray emitted within a certain solid angles induces current similar to that from charged particles emitted and seen by the chamber in typical RHIC collisions. For iTPC sector, all anode wires are divided into 4 groups, which are corresponding to 4 HV channels.  Figure~\ref{fig:StrongTest} shows the leakage current variation of all 4 channels. The leakage current of one channel reached about 500 nA with no trip and sparks observed. The decrease of leakage current versus time is found to be driven by the intensity-drop of X-ray source.

\section{Summary}

We report the details for the first iTPC MWPC prototype sector construction and testing at SDU, which is part of the STAR BES-II detector upgrades. Based on the measurements from a laser based system, the prototype has the wire tension and pitch precision better than 6$\%$ and 10 $\mu$m, respectively. The performance test of the iTPC MWPC prototype shows that the uniformity of the gas gain and the energy resolution of the prototype are better than 1$\%$ (RMS) and 20$\%$ (FWHM), respectively. In irradiation test under 230 kHz X-ray, a reasonable stability has been observed. All these results show that the prototype meets the iTPC upgrade requirements and the construction techniques are ready for mass production.

\section*{Acknowledgments}

We would like to thank the STAR iTPC upgrade collaboration mem- bers for their valuable suggestions and support. The work was supported in part by the National Natural Science Foundation of China (No. 11520101004), the Major State Basic Research Development Program in China (No. 2014CB845400), and the Natural Science Foundation of Shandong Province, China, under Grant No. ZR2013JQ001. This manuscript has been co-authored by employees of Brookhaven Science Associates, LLC, under Contract No. DE-SC0012704 with the U.S. De- partment of Energy.

%\section*{References}

\end{document}